\documentclass[english,preprint,aps,superscriptaddress,showkeys,showpacs,pre]{revtex4}
\usepackage[utf8x]{inputenc}
\usepackage{amsmath}
\usepackage{amssymb}
\usepackage{graphicx}
\usepackage{epstopdf}
\usepackage{setspace}

\makeatletter
\@ifundefined{textcolor}{}
{%
 \definecolor{BLACK}{gray}{0}
 \definecolor{WHITE}{gray}{1}
 \definecolor{RED}{rgb}{1,0,0}
 \definecolor{GREEN}{rgb}{0,1,0}
 \definecolor{BLUE}{rgb}{0,0,1}
 \definecolor{CYAN}{cmyk}{1,0,0,0}
 \definecolor{MAGENTA}{cmyk}{0,1,0,0}
 \definecolor{YELLOW}{cmyk}{0,0,1,0}
}

\usepackage[english]{babel}
\usepackage{subfigure}

\usepackage{amsfonts}\usepackage{textcomp}\usepackage{color}
\usepackage{array}

\makeatother

\usepackage{babel}
\begin{document}

\title{Synchronization-Induced Spike Termination in Networks of Bistable Neurons}

\author{Muhammet Uzuntarla}
\email[]{muzuntarla@yahoo.com}
\affiliation{Department of Biomedical Engineering, Bulent Ecevit University, Zonguldak, Turkey}

\author{Joaquin J. Torres}
\affiliation{Department of Electromagnetism and Physics of the Matter and Institute Carlos I for Theoretical and Computational Physics, University of Granada, Granada, E-18071 Spain}

\author{Ali Calim}
\affiliation{Department of Biomedical Engineering, Bulent Ecevit University, Zonguldak, Turkey}

\author{Ernest Barreto}
\affiliation{Department of Physics and Astronomy and The Krasnow Institute for Advanced Study, George Mason University, Fairfax, Virginia, USA}

\date{\today}
\begin{abstract}

We observe and study a self-organized phenomenon whereby the activity in a network of spiking neurons spontaneously terminates. We consider different types of populations, consisting of bistable model neurons connected electrically by gap junctions, or by either excitatory or inhibitory synapses, in a scale-free connection topology. We find that strongly synchronized population spiking events lead to complete cessation of activity in excitatory networks, but not in gap junction or inhibitory networks. We identify the underlying mechanism responsible for this phenomenon by examining the particular shape of the excitatory postsynaptic currents that arise in the neurons. We also examine the effects of the synaptic time constant, coupling strength, and channel noise on the occurrence of the phenomenon. 
\end{abstract}


\maketitle

\section{Introduction}

Synchronization is a collective phenomenon that occurs in systems of interacting units, and is ubiquitous in nature, society, and technology \cite{strogatzbook,strogatz2000}. Research on synchronization has been carried out for decades. Investigators have focused on describing this behavior in many different natural and artificial systems, examining its role in the functioning of these systems, and ascertaining the fundamental underlying mechanisms that give rise to it.

The presence of synchronization has been widely reported in many living organisms \cite{glassbook,glass2001,herbert2016,bajec2009,mirollo1990}. A prominent example is neural synchronization, which has been observed in neural systems of many different animals from invertebrates to mammals \cite{buzsaki2004,buzsakibook,kreiter1996,uhlhaas2006,winfreebook,gillette2004}, and which occurs in both normal and pathological states \cite{jiruska2013,schindler2007,rubchinsky2012,babiloni2016,cantero2005,tallon2004}. On the other hand, other studies have focused on the possible mechanisms at microscopic, mesoscopic, and macroscopic scales that can induce and control synchronization in neural structures \cite{buzsaki2004,gonzalez2014,bonifazi2009,zhou2003,chao2005,bartos2007}. In general, findings from experimental and theoretical studies suggest that the emergence of neuronal synchronization results from the interplay between the intrinsic properties of individual neurons, synaptic interaction dynamics, network organization features, and where relevant, external inputs.

In neural systems, synchronization is widely considered to be responsible for the origin of oscillatory brain rhythms which have been associated with many cognitive tasks. It has been shown that synchronization in the electrical activity of neuronal populations might play significant roles in processes ranging from simple sensory transmission to perception and attention as well as learning \cite{Engel2001, Schall2001, BuzsakiWatson2012, WatsonBuzsaki2015, Womelsdorf2007}. In addition, different temporal patterns of synchronization have been observed in pathological conditions such as Alzheimer’s disease, Parkinson’s disease, and epilepsy \cite{BABILONI2016, Lehnertz2009, Hammond2007}. Therefore, elucidating the underlying bases of neuronal synchronization and the emergent consequences is a critical step in understanding how neural systems work.

An interesting synchronization-induced emergent behavior in neural systems is the termination of ongoing population activity \cite{Gutkin2001, Gutkin2004, Dipoppa2012, Dipoppa2013, Gutkin16}. In this phenomenon, persistent sustained activity (which is considered to be the prevalent neural substrate of working memory) in a recurrently-connected excitatory/inhibitory balanced population can be turned off by delivering a sufficiently strong external stimulus to the excitatory neurons in the population. This causes those neurons to spike synchronously, and subsequently, they remain silent. Mechanisms that prevent synchronization-induced spike termination (SIST), and thus serve to maintain sustained activity, have been explored in a system of two coupled quadratic integrate-and-fire neuron models \cite{Gutkin16}. In a recent work \cite{Uzuntarla2017,Calim2017}, we found similar spiking activity termination in a complex network of excitatory bistable Hodgkin-Huxley (HH) neurons. In particular, we found that neural activity in an excitatory population stops suddenly when the coupling between neurons is strong enough to induce hypersynchronization in the network. Moreover, this happens spontaneously, without the need for any external input. This indicates the existence of an internal mechanism responsible for neural activity termination which is triggered by the synchronized activity in the population. However, the precise mechanisms underlying SIST have not been fully characterized. In addition to its possible role in controlling sustained activity in working memory tasks, SIST could also be involved in the spontaneous termination of epileptic seizures, either in humans or animals, or seizure-like events observed in \emph{in vitro} preparations. In these situations, epileptic seizures also typically terminate spontaneously due to the emergence of hypersynchronous states in specific brain regions \cite{Schindler2007, Timofev2004, Lado2008}.

Due to these possible connections, it is interesting to investigate the SIST phenomenon in more detail by considering cellular and synaptic dynamics in different types of populations present in the brain. Here, we study the behavior of three types of networks composed of bistable HH neurons with complex network topology, involving either electrical gap junctions or chemical synapses that are either excitatory or inhibitory. We find that periodic synchronous population activity emerges in all three networks, and the spontaneous termination of activity only appears in the network with excitatory synapses. We show that this characteristic feature of excitatory populations arises from the particular shape of the synaptic current profile that arises in the presence of strong population synchronization. We also investigate the effect of different parameters that control various synaptic current features on the emergence of collective spike termination.

\section{Models and Methods}

A single neuron of our networks is modeled based on the Hodgkin-Huxley (HH) equations as follows \cite{HH1952b}:                                                                                   
\begin{equation}
C_m \frac{dV_i}{dt}= -G_{Na}(m_i,h_i)(V_i-E_{Na})-G_K(n_i)(V_i-E_K)-G_L(V_i-E_L)+I_0+I_i^{syn},
\end{equation}
where $C_m=1\mu F/cm^{2}$ is the capacitance of the cell membrane per unit area and $V_i$ is the membrane potential of the $i$-th neuron in millivolts. $I_{0}$ is a constant external bias current in $\mu A/cm^{2}$ used to adjust neuronal excitability. $E_{Na}=115 mV$, $E_K=−12 mV$, and $E_L=10.6 mV$ are the reversal potentials for the sodium, potassium, and leak currents, respectively, and $G_{Na}$, $G_{K}$, and $G_{L}$ are the corresponding channel conductances. The leak conductance is assumed to be constant, with $G_L=0.3 mS/cm^{2}$, while the sodium and potassium conductances change dynamically according to
\begin{eqnarray}
G_{Na}(m_i,h_i)= g_{Na}^{max}m_{i}^{3}h_{i} \label{eq1}\\
G_{K}(n_i)= g_{K}^{max}n_{i}^{4} \label{eq2}.
\end{eqnarray}
Here, $g_{Na}^{max}=120 mS/cm^{2}$ and $g_{K}^{max}=36 mS/cm^{2}$ are the maximal sodium and potassium conductances. The product $m_i^3h_i$ is the mean proportion of open sodium channels in cell $i$, where $m_i$ and $h_i$ are the activation and inactivation gating variables. Likewise, $n_i^4$ is the mean proportion of open potassium channels in cell $i$, with $n_i$ being the corresponding activation gating variable. 

To incorporate stochastic dynamics for the ion channels, we use the well-known Fox algorithm due to its widespread use and computational efficiency \cite{fox97}. In Fox’s algorithm, the gating variables obey the Langevin equation
\begin{equation}
\frac{dx_i}{dt}=\alpha_{x_i}(1-x_i)-\beta_{x_i}{x_i}+\xi_{x_i}(t),
\end{equation}
where $\alpha_{x_i}$ and $\beta_{x_i}$ are the rate functions for the gating variables ${x_i}=m_i,n_i,h_i$ and are given by \cite{pankratova2005b}
\begin{eqnarray}
\alpha_{m_i}=0.1\frac{(25-V_i)}{ exp[(25-V_i)/10]-1}\label{eq3}\\
\beta_{m_i}=4 exp[-V_i/18]\label{eq4}\\
\alpha_{n_i}=0.01\frac{(10-V_i)}{ exp[(10-V_i)/10]-1}\label{eq5}\\
\beta_{n_i}=0.125 exp[-V_i/80]\label{eq6}\\
\alpha_{h_i}=0.07 exp[-V_i/20]\label{eq7}\\
{\beta_{h_i}}=\frac{1}{ exp[(30-V_i)/10]+1}\label{eq8}.
\end{eqnarray}

The stochastic nature of the channels is described via the independent zero-mean Gaussian white noise sources $\xi_{x_i}(t)$, whose autocorrelation functions satisfy following equations \cite{fox97}:
\begin{eqnarray}
\langle \xi_{m_i}(t)\xi_{m_i}(t') \rangle=\frac{2\alpha_{m_i}\beta_{m_i}}{ N_{Na}(\alpha_{m_i}+\beta_{m_i})}\delta(t-t') \label{eq9}\\
\langle \xi_{h_i}(t)\xi_{h_i}(t') \rangle=\frac{2\alpha_{h_i}\beta_{h_i}}{ N_{Na}(\alpha_{h_i}+\beta_{h_i})}\delta(t-t') \label{eq10}\\
\langle \xi_{n_i}(t)\xi_{n_i}(t') \rangle=\frac{2\alpha_{n_i}\beta_{n_i}}{ N_{K}(\alpha_{n_i}+\beta_{n_i})}\delta(t-t') \label{eq11},
\end{eqnarray} 
where $N_{Na}$ and $N_{K}$ represent the overall numbers of the sodium and potassium ion channels within the membrane patch. Assuming homogenous sodium and potassium ion channel densities $\rho_{Na}= 60\mu m^{-2}$ and $\rho_{K}= 18\mu m^{-2}$, the total ion channel numbers are given by $N_{Na}= A\rho_{Na}$ and $N_{K}= A\rho_{K}$. Thus $A$, the membrane patch area, globally determines the intrinsic noise level \cite{White2000, Schmid2003, Ozer2009JTB}. Unless otherwise stated, we use $A=10^5\mu m^{2}$. Note that the $\alpha_{x_i}$ and $\beta_{x_i}$ in the autocorrelation formulas (Eqs.~\ref{eq9}-\ref{eq11}) have instantaneous values for the time step involved \cite{fox97}.
Finally, $I_i^{syn}$ in Eq (1) is the total synaptic current received by neuron $i$. We consider coupling via electrical gap junctions and chemical synapses separately. In the case of gap junctions, the synaptic current is modeled with linear electrical coupling, summing over neighbors:
\begin{equation}
I_i^{syn}=\sum_{j \in neighbors(i)} g_{syn}(V_{ j}-V_{ i}).
\end{equation}
where $neighbors(i)$ is the set of neighbors of neuron $i$. For chemical synapses, the synaptic current takes the form \cite{kochbook}
\begin{eqnarray}
I_i^{syn}= \sum\nolimits_{j \in neighbors(i)} g_{syn}s_{ j}(E_{rev}-V_{i})\label{eq12}\\
\dot s_j=-s_j/\tau_{syn}+\delta(t-t'_j) \label{eq13},
\end{eqnarray}
where $s_{ j}$ is the fraction of open receptor channels for neuron $j$. Accordingly, once neuron $j$ emits a spike, $s_{ j}$ is updated with $s_{ j}\leftarrow s_{ j}+1$. Then, it decays exponentially with time constant $\tau_{syn}$. Unless stated otherwise, we use $\tau_{syn}=3ms$. The maximal conductance $g_{syn}$ will be a parameter of interest in our investigations. The parameter $E_{rev}$ is the synaptic reversal potential. We use $E_{rev}=70mV$ for excitatory synapses and $E_{rev}=-10mV$ for inhibitory ones.

We obtain bistability in each neuron's dynamics by setting the external input current $I_0$ to a value between those that correspond to the creation of stable and unstable limit cycles by saddle-node bifurcation (at $I_{0}=6.26 \mu A/cm^2$) and the subcritical Hopf bifurcation (at $I_{0}=9.78 \mu A/cm^2$), where the aforementioned unstable limit cycle coalesces with the stable resting equilibrium. Specifically, we set $I_{0}=6.80 \mu A/cm^2$, for which the stable limit cycle and the resting equilibrium have good-sized basins of attraction.

To connect the neurons, we use the preferential attachment algorithm to generate a scale-free (SF) network \cite{Barabassi99, Albert2002, Heuvel2010}. The construction process begins with a set of $m$ fully-connected nodes, and subsequently every new node is attached to $m$ different nodes already present in the network. In the preferential attachment algorithm, the probability $\Pi $ that a new node connects to node $i$ depends on that node's degree $k_i$ according to $\Pi =k_i/\sum_j k_j$. After many iterations, this procedure yields a network with average degree $\langle k \rangle =2m$ and power-law degree distribution with exponent $-3$. We use $m=10$ and $N=200$ nodes throughout this work. We found that this network size was sufficient for our purposes, as simulations with larger networks (not shown) exhibit the same qualitative features that we investigate in this paper.

Finally, to characterize the population spiking behavior quantitatively, we calculate the mean firing rate of the network, averaged over trials, as follows. For each trial, initial conditions are randomly and independently selected for all neurons with uniform probability from $−10$ to $80mV$ for the membrane voltage variable $V_i$, and from $0$ to $1$ for each of the gating variables $m_i$, $n_i$, and $h_i$. The system equations are then integrated for a time $T = 1 s$ to eliminate transients. Then, over the following $\tau = 5 s$, we count the number of spikes $S_{ij}$ generated by the $i$th neuron during trial $j$. We define a spike as an upward crossing of the membrane potential through $20 mV$. The whole procedure is then repeated $L=20$ times, and the population mean firing rate is

\begin{equation}
\nu=\frac{1}{LN\tau} \sum_{j=1}^{L} \sum_{i=1}^{N} S_{ij}.
\end{equation}

We numerically integrate using the Euler-Maruyama algorithm with a step size of $10 \mu s$.

\section{Results}

We investigate the firing behavior of bistable Hodgkin-Huxley neurons interacting via a scale-free network. We first consider three separate neuron populations internally coupled by (1) electrical connections (i.e.~gap junctions), (2) excitatory chemical synapses, and (3) inhibitory chemical synapses. For ease of exposition, we will refer to these networks as the electrical, excitatory, and inhibitory networks (or populations), respectively. We will also often refer to the maximal conductance $g_{syn}$ of these connections as the coupling strength. Subsequently, we will concentrate on the behavior of the excitatory network.

\begin{figure}
\begin{centering}
\includegraphics[scale=.6]{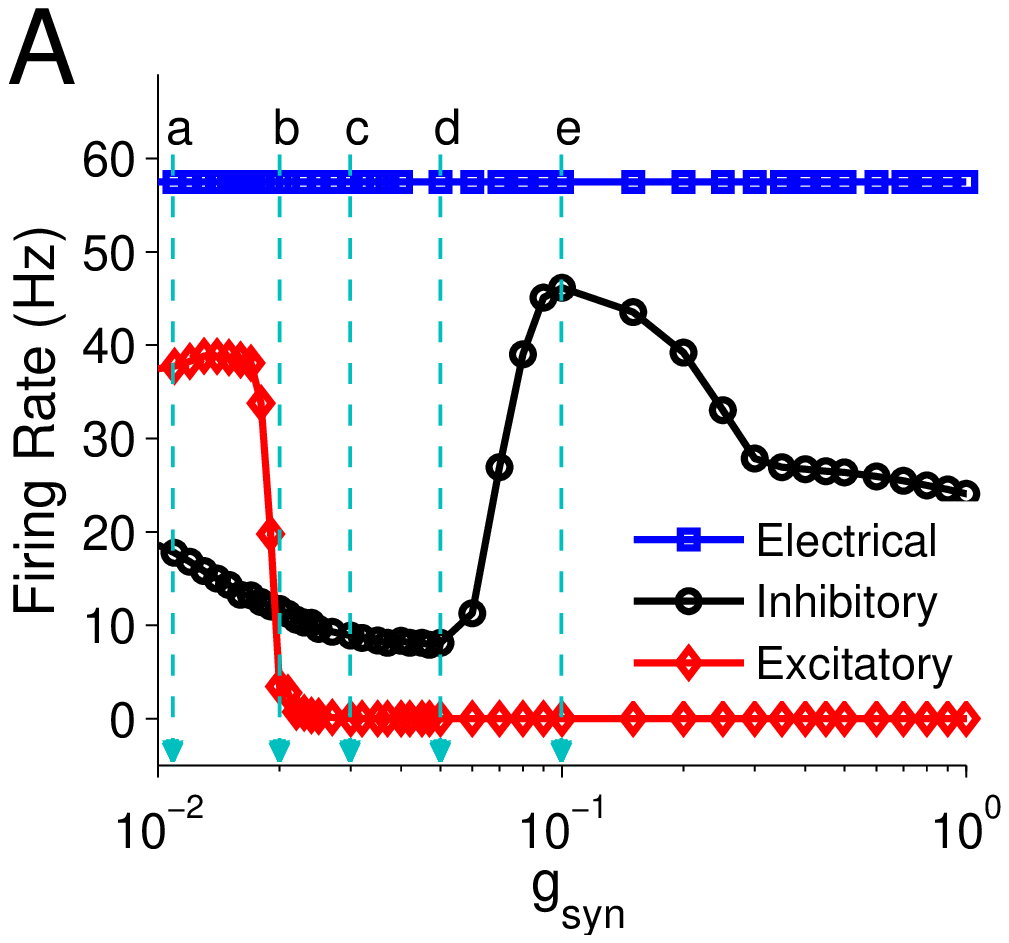}
\includegraphics[scale=.6]{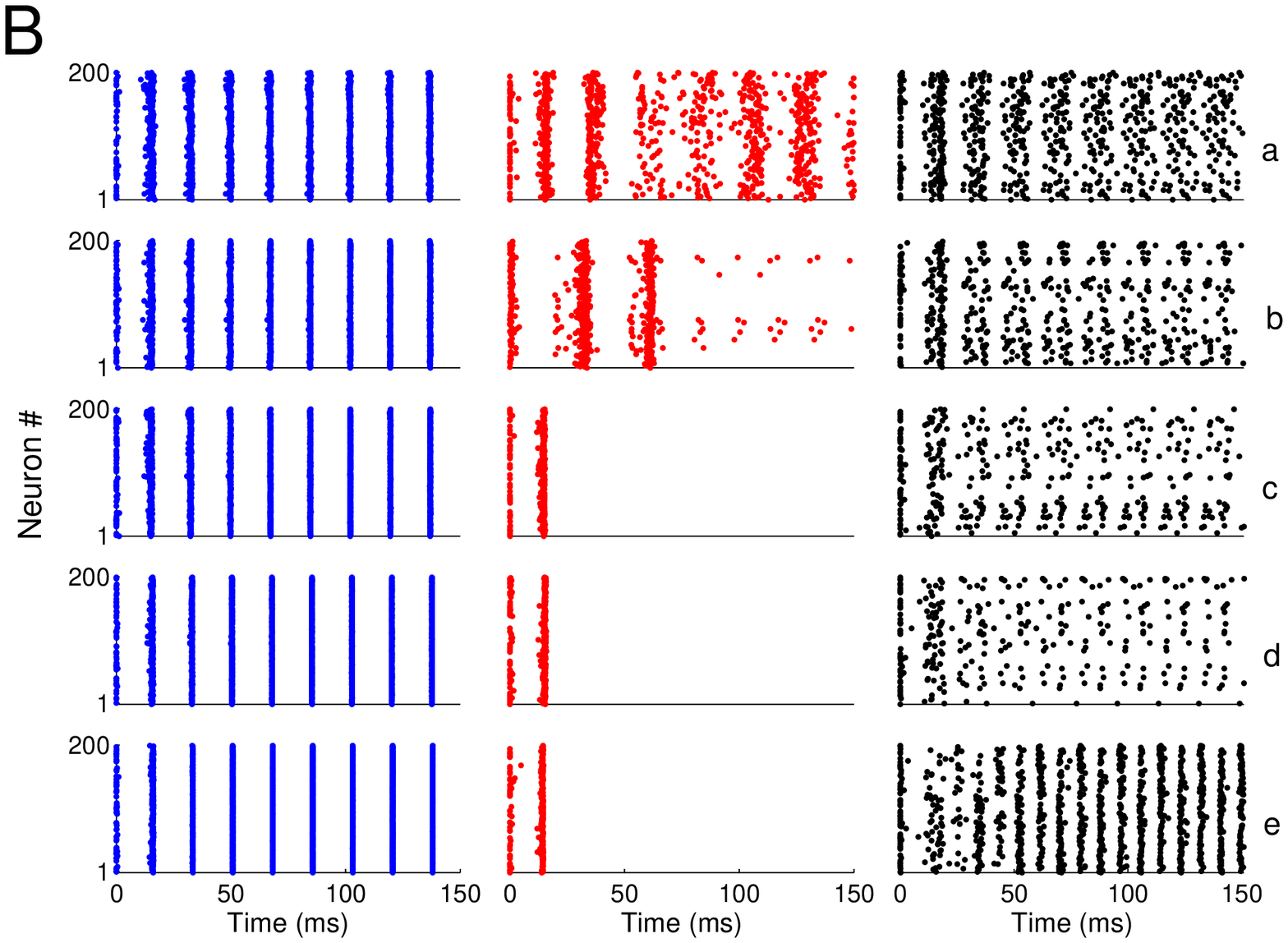}
\par\end{centering}
\caption{\setstretch{1.0}\label{scats}The firing behavior of different neural populations. A) The population mean firing rate versus the coupling strength $g_{syn}$ in the electrical (blue), excitatory (red), and inhibitory (black) networks. B) Raster plots of the activity in the same networks, arranged in columns left to right, respectively. The rows, labeled a-e (on the right), correspond to the $g_{syn}$ values similarly marked in panel A, where for $a$, $g_{syn}=0.01$; $b$, $g_{syn}=0.02$; $c$, $g_{syn}=0.03$; $d$, $g_{syn}=0.05$; and $e$, $g_{syn}=0.1$.}
\end{figure}

Fig.~\ref{scats}A shows the population mean firing rate versus the coupling strength for each network. 
The three populations exhibit quite different behavior as the strength of the coupling increases. We see that in the electrical network, increasing the conductance does not change the mean firing rate at all. However, in the excitatory and inhibitory networks, increasing the conductance leads to very significant and different changes of the population mean firing rate. In the inhibitory network, increasing the coupling strength induces non-monotonic changes in the mean firing rate: first it decreases, then increases to a maximum, and then decreases again, ultimately saturating at an intermediate value. In contrast, in the excitatory network, there is a dramatic dropoff of activity near a particular value of the synaptic conductance, and no activity for higher coupling strengths.

To better understand these results, we made raster plots of each population at the various notable values of $g_{syn}$ marked (a)-(e) in Fig.~\ref{scats}A. These are shown in Fig.~\ref{scats}B for the electrical, excitatory, and inhibitory networks, arranged in columns left to right, respectively. Plots for increasing $g_{syn}$ values are arranged from top to bottom. It is immediately obvious that in the electrical network (left column), the population spiking activity is highly synchronous and periodic. The raster plots in this column are essentially the same, but a slight increase in the speed at which strong synchronization is achieved is visible as $g_{syn}$ increases.

In the case of the inhibitory network (right column), the activity is at best only weakly synchronized and weakly periodic in plots (a)-(d). We see that as the coupling strength increases, different patterns appear and the spiking becomes more sparse, consistent with the curve in Fig.~\ref{scats}A. That curve increases sharply after the $g_{syn}$ value marked (d), and we see in the last raster plot, panel B(e) for $g_{syn}=0.1$, that a strongly synchronous periodic pattern emerges with a burst frequency twice that of the electrical case. In fact, the population splits into two synchronized subpopulations that alternately fire as in the ``ING” mechanism for gamma rhythm generation \cite{Whittington1995, Wang1992, Kopell2004}.

Finally, in the excitatory population (middle column), synchrony initially increases quickly as the coupling increases. In panel (b), the activity abruptly becomes very sparse after the last strong population spike. In panels (c)-(e), all activity ceases after a highly synchronous spiking event that occurs very early in the simulation. This termination of spiking activity, and how it happens, is our main focus in the remainder of this work.

\begin{figure}
\begin{centering}
\includegraphics[width=15cm]{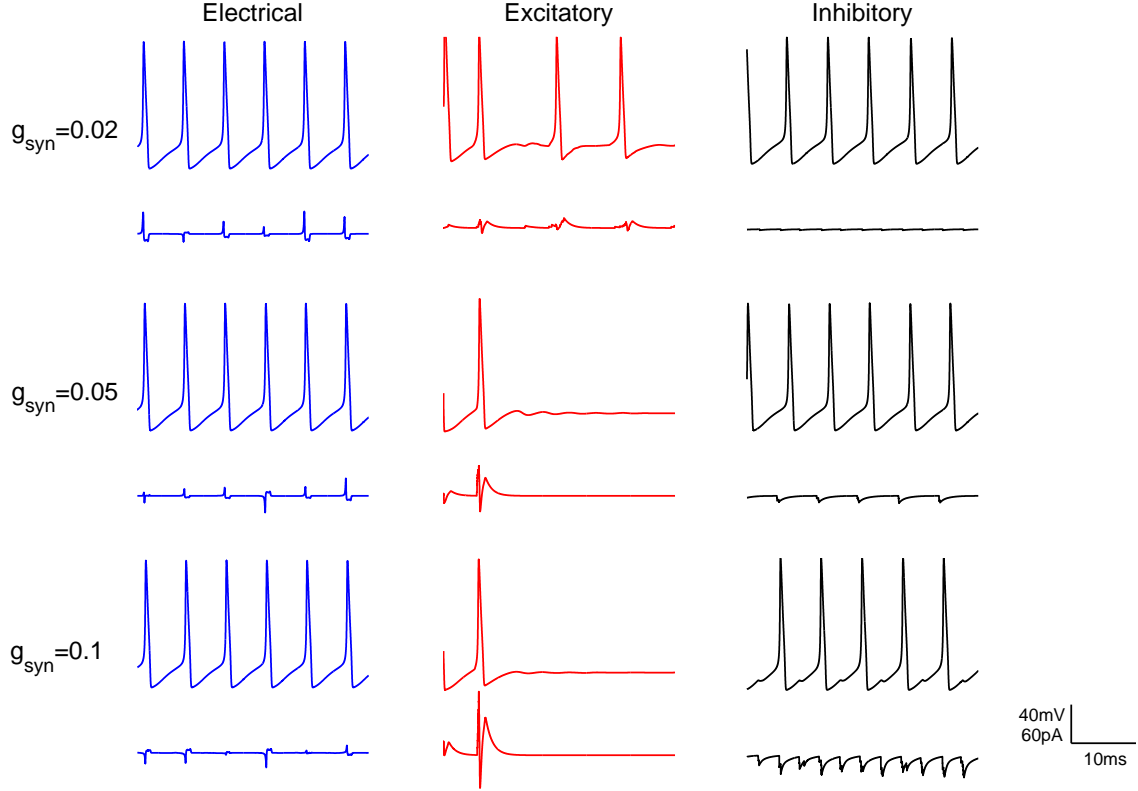}
\par\end{centering}
\caption{\setstretch{1.0}\label{terminationdemo}Membrane voltage and synaptic current traces in a randomly chosen neuron from the electrical, excitatory, and inhibitory networks, with coupling strengths $g_{syn}$ as shown.}
\end{figure}

To understand this spike termination effect, we examined the synaptic currents received by a single representative neuron drawn from each of the three population types. This is shown in Fig.~\ref{terminationdemo}, where membrane potential traces are plotted together with the corresponding total synaptic current inputs, for three values of the coupling strength (b, d, and e as in Fig.~\ref{scats}B, that is, $g_{syn}=0.02$, $0.05$, and $0.1$). First, we observe no significant change in the firing behavior in the electrical and inhibitory populations, except the small kinks that occur out of phase with the spiking in the strongly-coupled inhibitory case. In contrast, the neuron from the excitatory population stops spiking. Also, there is a dramatic change in the synaptic current amplitude as $g_{syn}$ increases in the case of the excitatory population. Note that the shapes of the synaptic currents have different temporal profiles in the different networks. In the excitatory case, the synaptic current is larger and exhibits a triphasic structure that we identify as follows. First, there is a sharp positive deflection and a quick return to the baseline. This is immediately followed by a sharp negative deflection and another quick return to the baseline. Finally, a less abrupt positive ``bump'' occurs in the current, after which it returns to baseline relatively slowly. The shape of this current trace arises from the competition between the membrane voltage and the excitatory synaptic reversal potential. We hypothesize that among these features are the cause of the spike termination phenomenon of interest.

To test the importance of these three phases of the synaptic current profile on spike termination, we perform an artificial stimulation protocol as follows: First, we record a synaptic current trace that induces network-wide spike termination in the excitatory network with $g_{syn}=0.05$. We then manipulate each phase of the synaptic current trace separately, inject it into a single isolated neuron, and observe its behavior. 

\begin{figure}
\begin{centering}
\includegraphics{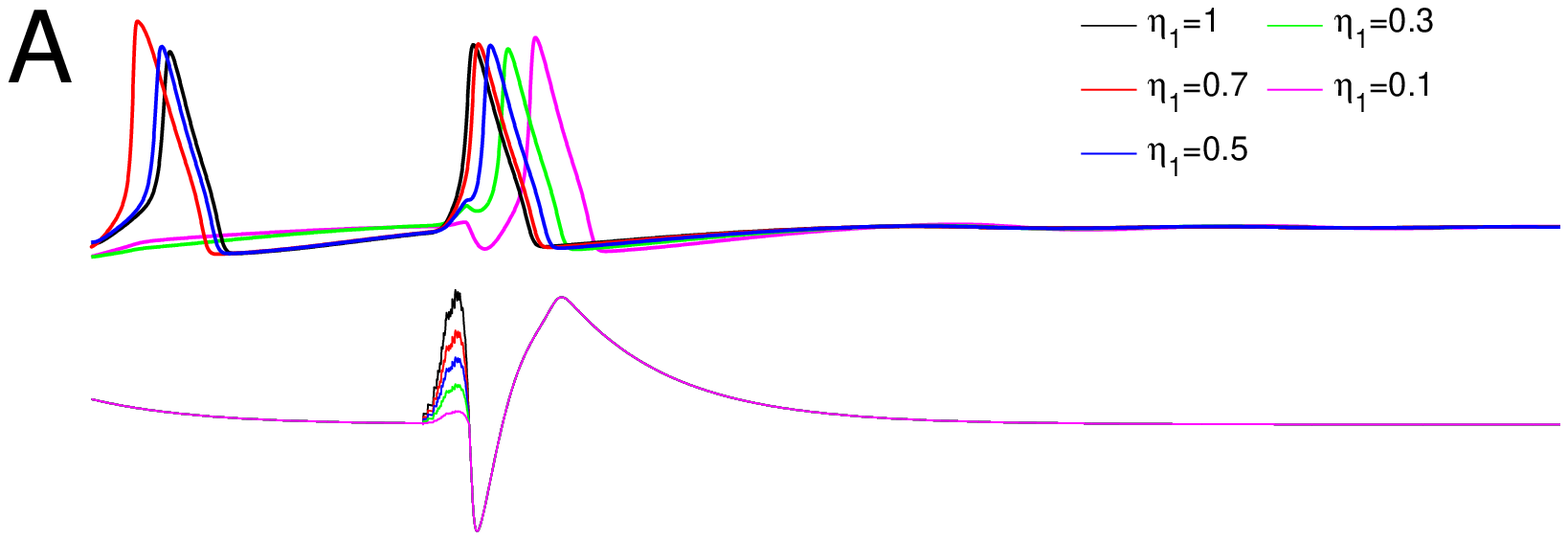}\\
\includegraphics{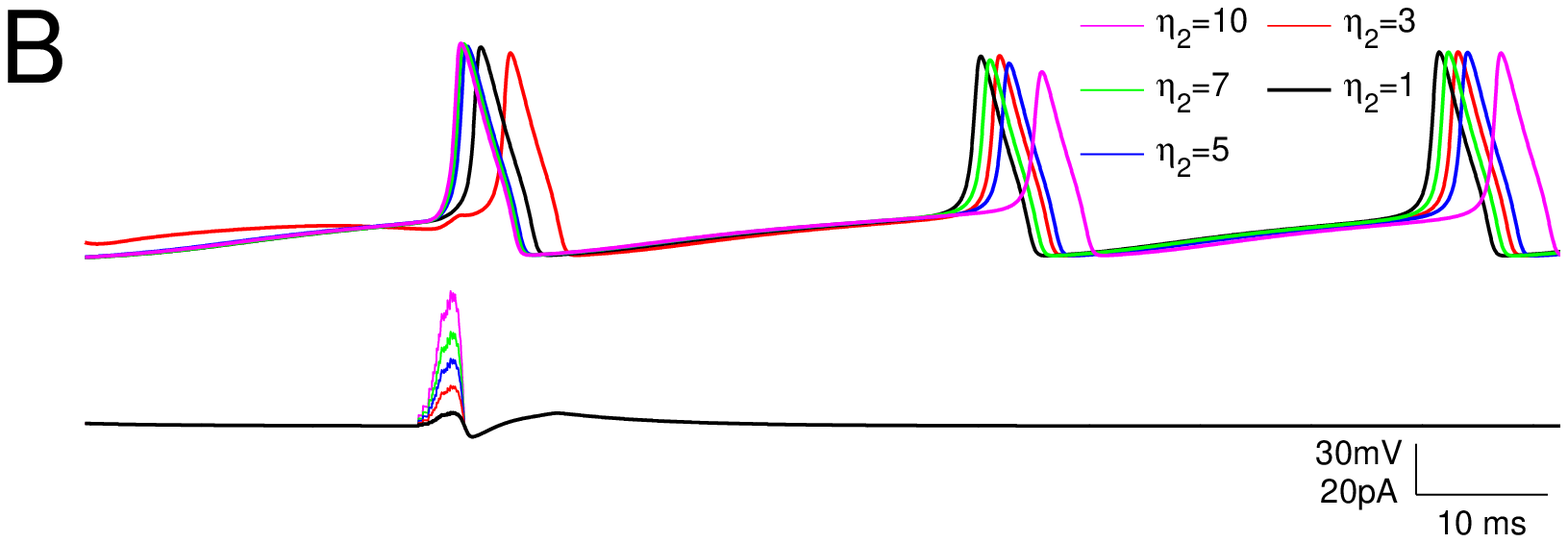}
\par\end{centering} 
\caption{\setstretch{1.0}\label{isynfp}The influence of the first peak of the synaptic current profile on spike termination. The top panels show voltage traces of the neuron when subjected to the modified versions of the frozen synaptic current trace shown in the bottom panels. In (A), the initial depolarizing phase is successively attenuated by factors $\eta_1$. In (B), the entire frozen synaptic current trace is first attenuated by $90\%$ to remove the spike termination effect, and then the depolarizing first phase is successively amplified by factors $\eta_2$.}
\end{figure}

\begin{figure}
\begin{centering}
\includegraphics{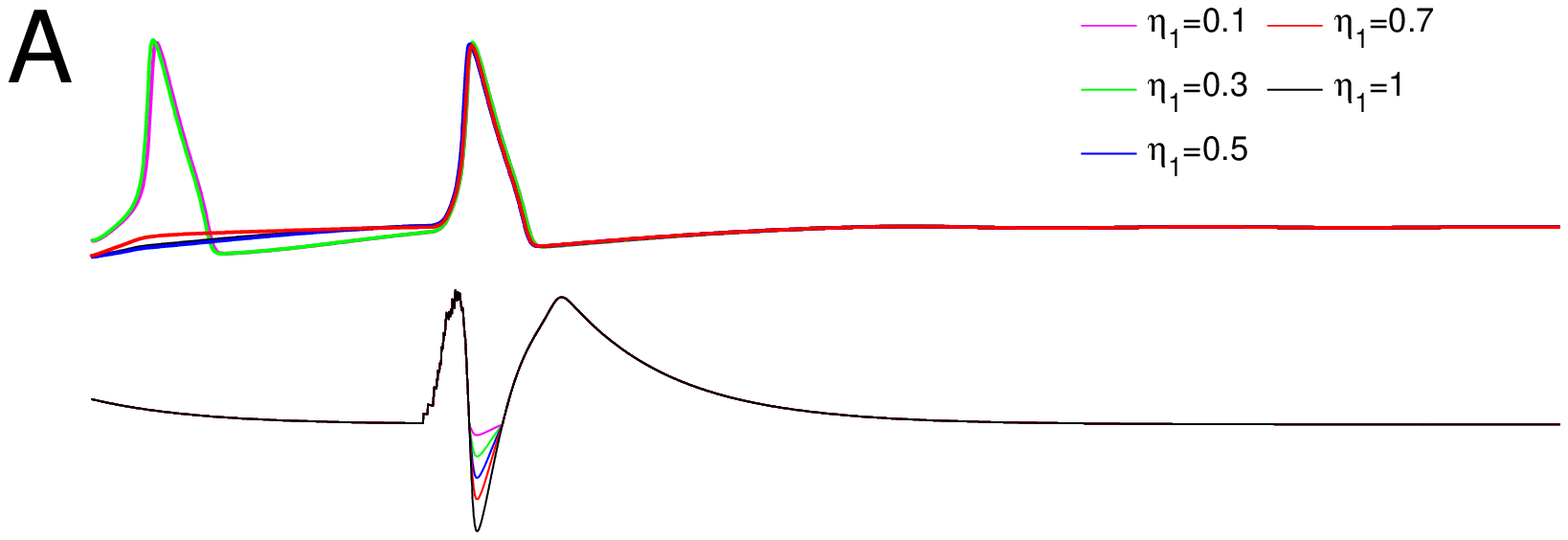} \\
\includegraphics{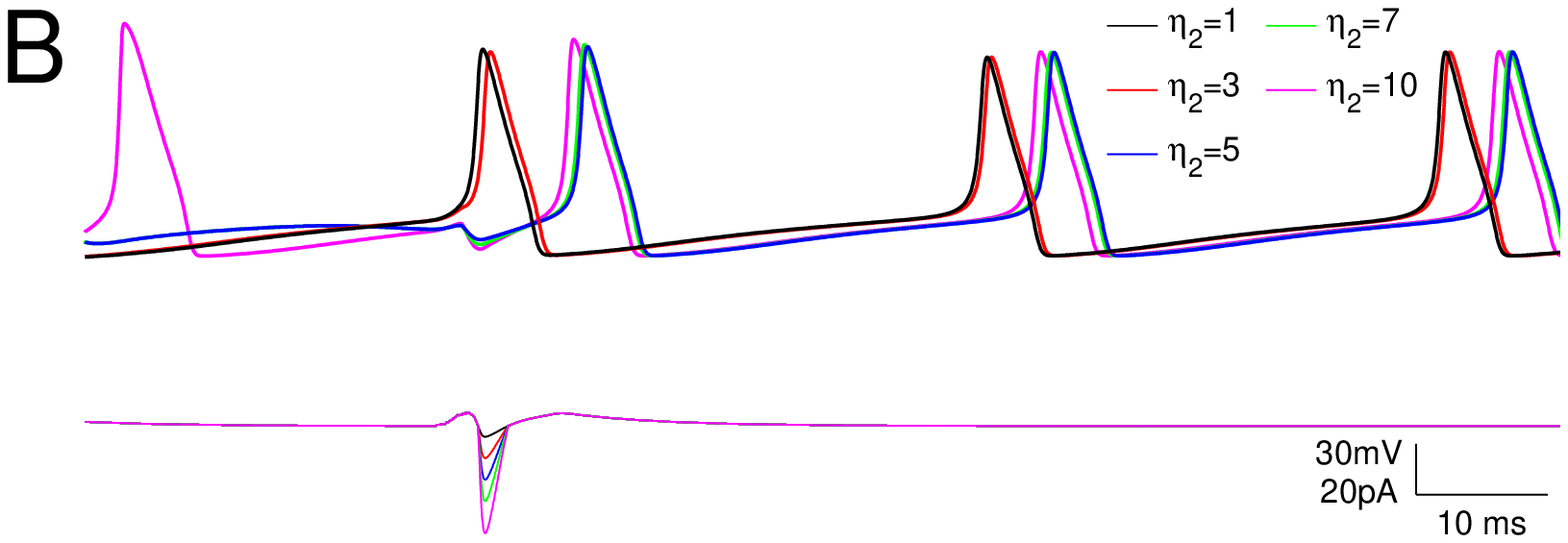} 
\par\end{centering}
\caption{\setstretch{1.0}\label{fig:minimum}The influence of hyperpolarizing phase of the synaptic current profile on spike termination. The top panels show voltage traces of the neuron when subjected to the modified versions of the frozen synaptic current trace shown in the bottom panels. The modifications are made as in Figure \ref{isynfp}: In (A), the second phase is attenuated by factors $\eta_1$ as shown. In (B), the entire frozen synaptic current trace is first attenuated by $90\%$ to remove the spike termination effect, and then the hyperpolarizing phase is successively amplified by factors $\eta_2$.}
\end{figure}

\begin{figure}
\begin{centering}
\includegraphics{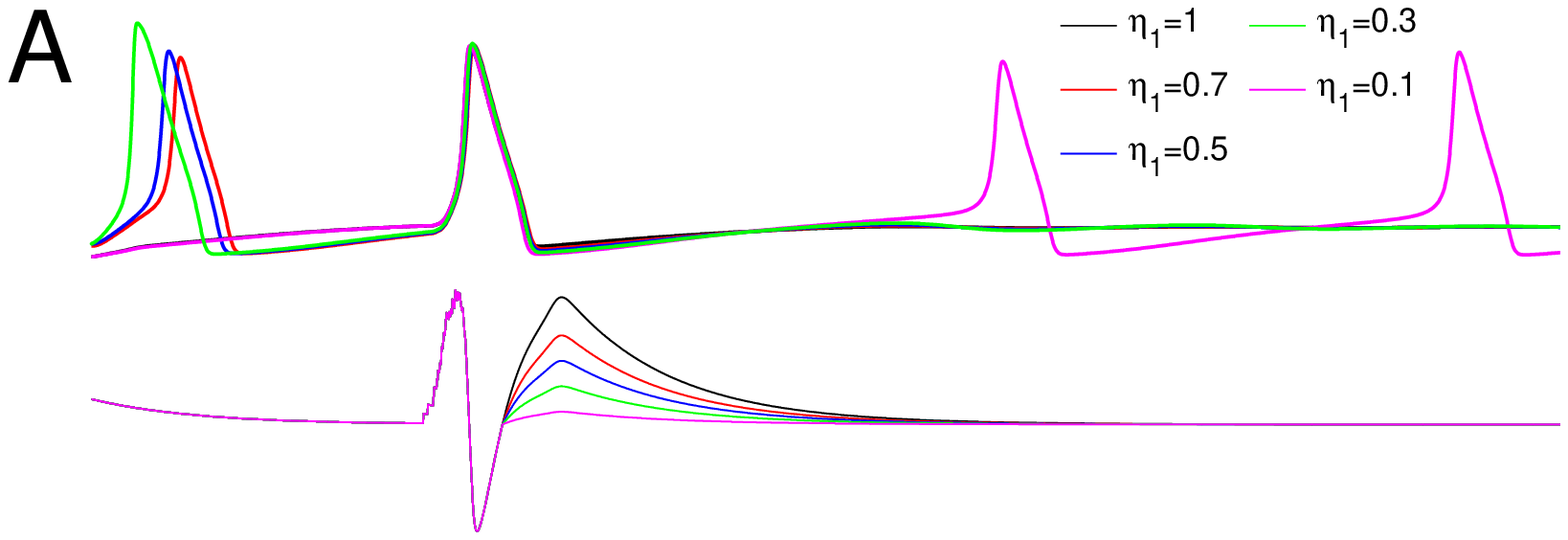} \\
\includegraphics{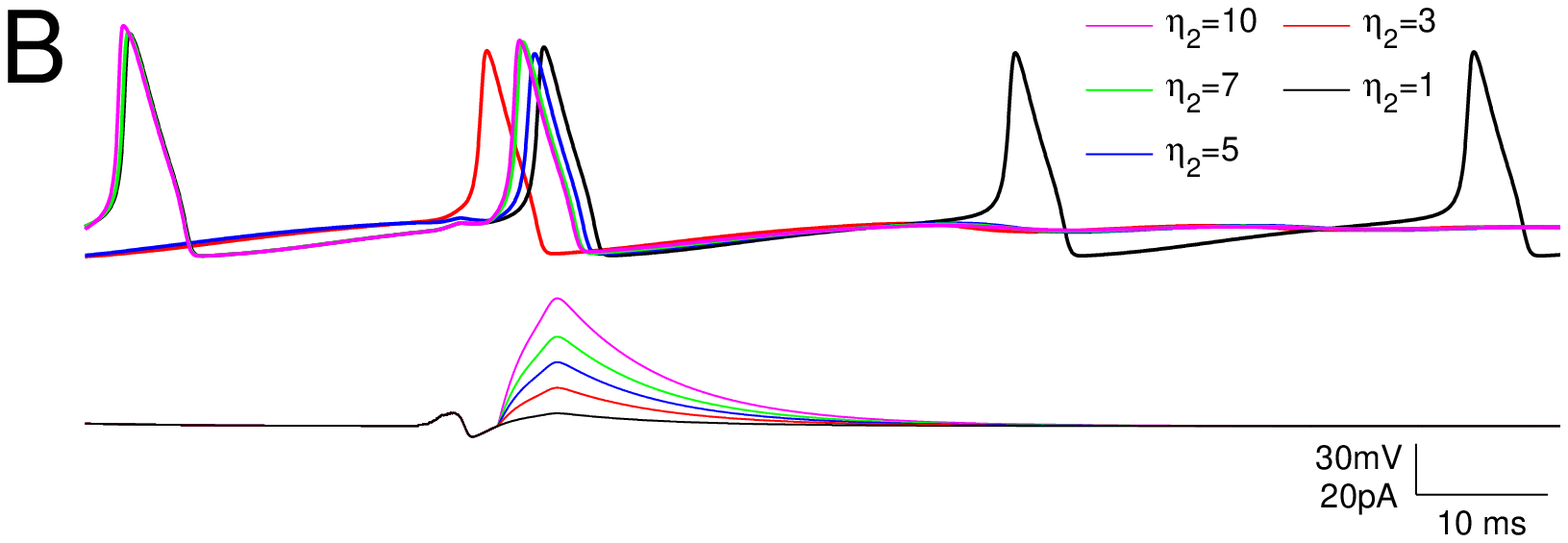}
\par\end{centering}
\caption{\setstretch{1.0}\label{secondpeak}The influence of the third phase of the synaptic current profile on spike termination. The top panels show voltage traces of the neuron when subjected to the modified versions of the frozen synaptic current trace shown in the bottom panels. The modifications are made as in Figure \ref{isynfp}: In (A), the third phase is attenuated by factors $\eta_1$ as shown. The termination effect, present for $\eta_1=1$ through $0.3$, disappears and spiking resumes for $\eta_1=0.1$. In (B), the entire frozen synaptic current trace is first attenuated by $90\%$ to remove the spike termination effect, and then the slow depolarizing phase is successively amplified by factors $\eta_2$. The termination effect returns for $\eta_2 \geq 3$.}
\end{figure}

The results are shown in Figs.~\ref{isynfp}-\ref{secondpeak}. (Note that in each case, the single isolated neuron is initiated with random initial conditions. This leads to some scatter in the membrane voltage traces, but what is important is whether or not the spiking continues or terminates following the injection of the modified synaptic current.)

To show the influence of the first phase of the synaptic current, we attenuate just that portion by various values of a factor $\eta_{1} \leq 1$, and leave the other two phases unaltered. This is shown with different colored solid lines in the bottom panel of Fig.~\ref{isynfp}A.
The corresponding neuronal responses are shown in the color-matched voltage traces in the top panel. We find that for all values of $\eta_1$ tested, the injected current trace does not change the spike termination effect, as is seen in the case with $\eta_{1}=1$. We then attenuate the entire frozen synaptic current trace by $90\%$ in order to remove the spike termination effect, and then progressively amplify just the first phase by factors $\eta_{2} \geq 1$. The results are shown in Fig.~\ref{isynfp}B. We see that for all values of $\eta_2$ tested, the neuron spikes regularly after the synaptic current profile has passed. Thus, we conclude that the depolarizing first phase of the synaptic current plays no role in the spike termination effect in the network.

In Fig.~\ref{fig:minimum}, the same protocol is followed to investigate the influence of the hyperpolarizing second phase of the synaptic current profile.
Again, we find that for all values of $\eta_{1}$ and $\eta_{2}$ tested, the initial behavior of the postsynaptic neuron did not change. Thus, we conclude that the hyperpolarizing second phase of the synaptic current also plays no role in the spike termination effect.

The same analysis applied to the depolarizing third phase of the current profile leads to different results, as shown in Fig.~\ref{secondpeak}.
As before, the endogenous synaptic current (i.e., $\eta_1 = 1 = \eta_2$) leads to spike termination. With increasing attenuation of the third phase by the increasing values of the factor $\eta_1$ shown in Fig.~\ref{secondpeak}A, termination remains until $\eta_1 = 0.1$ is reached. For this value, the neuron spikes regularly after the synaptic current event passes. Thus, for values of $\eta_1$ below a critical value (somewhere between $0.3$ and $0.1$), the spike termination effect disappears. Attenuating the entire current trace by $90\%$ removes the spike termination effect as before, but as is shown in Panel B, it returns (i.e., spiking ceases) when the attenuated third phase is then amplified by a factor of $\eta_2 \geq 3$. These findings indicate that the amplitude of the third phase of the synaptic current input determines whether or not the spike termination effect occurs.

\begin{figure}
\begin{centering}
\includegraphics[width=8cm]{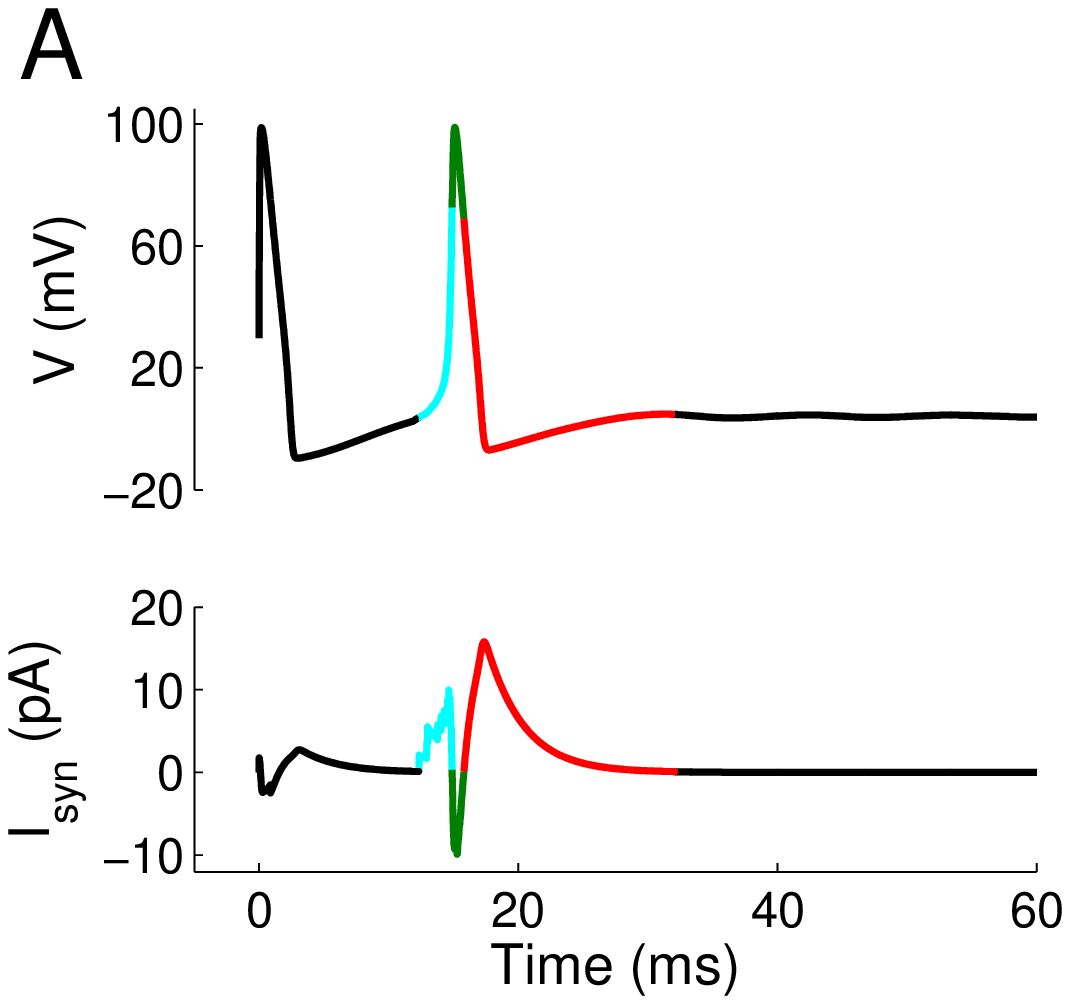}
\includegraphics[width=8cm]{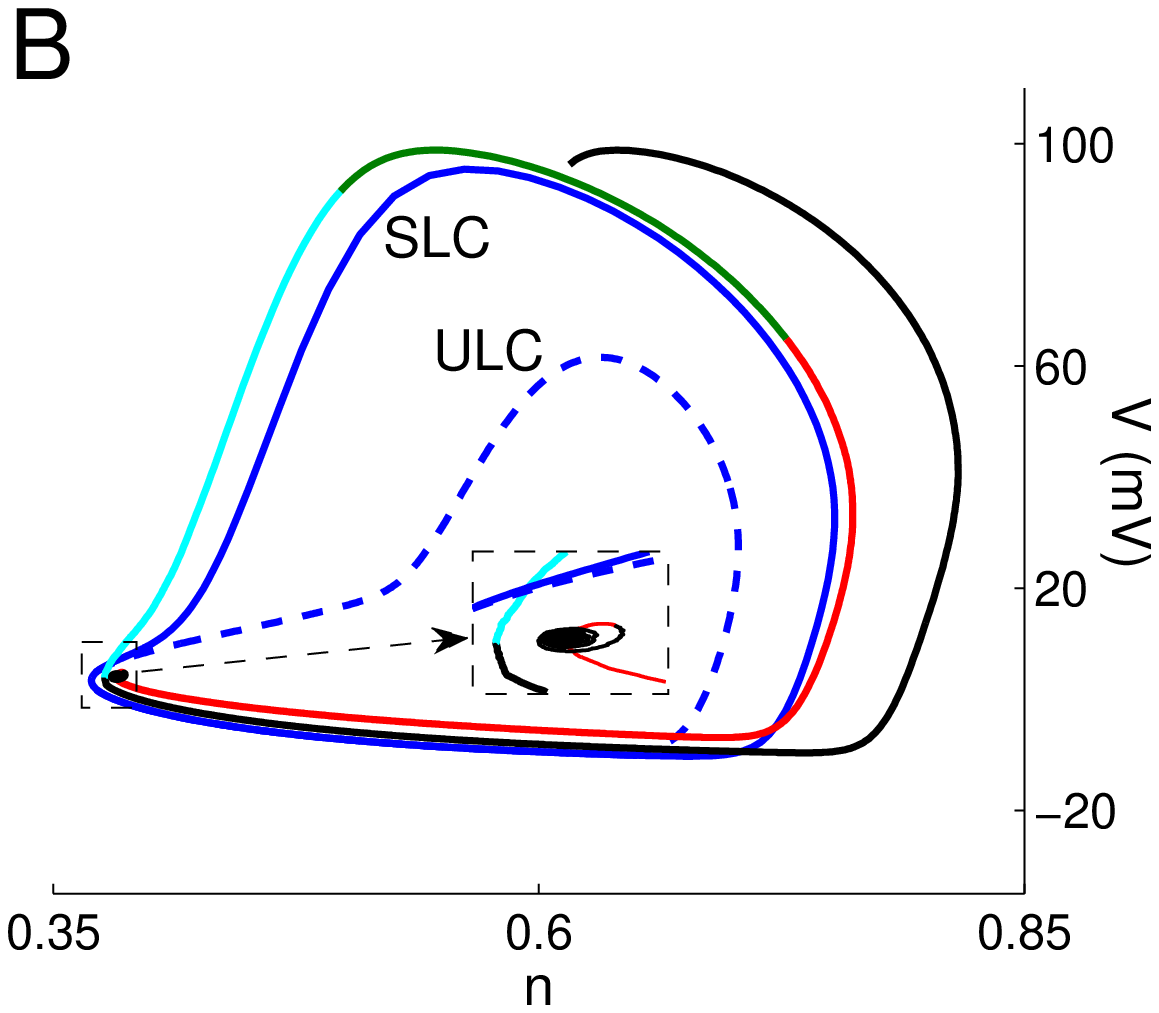}
\end{centering}
\caption{\setstretch{1.0}\label{phaseplot}Dynamical interpretation of a trajectory leading to spike termination. (A) The different phases of the membrane voltage (top) and synaptic current (bottom) are colored separately.  (B) Using data from the same trajectory, and using the same color scheme, the membrane voltage $V$ is plotted versus the gating variable $n$. The blue curves are the stable (SLC) and unstable (ULC) limit cycles. The dashed boxes show a magnification of the ``nose'', inside of which is found the stable equilibrium. Spiking terminates when the trajectory is trapped here.}
\end{figure}

We can identify the dynamical mechanism underlying spike termination by plotting the trajectory of the neuron's dynamics as projected onto the $V$-$n$ plane, and color-coding the time course of the data so that we can visually match the three phases of the postsynaptic current with the behavior of the membrane voltage. This is shown in Fig.~\ref{phaseplot}.

Fig.~\ref{phaseplot}A shows the membrane voltage and the synaptic current, plotted versus time, with the three phases of interest identified by color. Fig.~\ref{phaseplot}B shows data from the same trajectory projected onto the $V$-$n$ plane using the same color scheme. Also included, in blue and for reference, is the stable limit cycle (SLC; solid curve) and the unstable limit cycle (ULC; dotted curve). The SLC corresponds to autonomous spiking: during an action potential, the loop is traversed in a clockwise manner. Note also that the resting equilibrium sits inside the ``nose'' at the lower left portion of the SLC (see the magnification box). The boundary that separates the basins of the stable equilibrium and the SLC is mediated by the ULC. (More precisely, the basin boundary is the closure of the stable manifold of the ULC.)

The beginning of the trajectory, in black, shows an action potential already in progress, and the subsequent refractory period that follows. In panel B, the corresponding trace (also in black) begins in the upper middle region, swings down to the lower right, and then relaxes towards the ``nose''. The first phase of the synaptic current, colored in cyan, occurs as the neuron receives several inputs whose effects are seen as the jagged steps in the synaptic current trace. A new action potential is triggered, and the voltage increases quickly. Note in panel B that the trajectory begins its excursion deflected towards higher voltages relative to the SLC. Once the second phase begins, the trace changes to green, and the voltage achieves its maximum and drops abruptly. The corresponding trace in panel B shows the trajectory rounding the upper part of the SLC loop. The trajectory then changes to red, indicating the third phase. We see the downstroke of the action potential, and we infer from the $V$-$n$ phase plot that the perturbation to the trajectory due to the third-phase synaptic current causes the trajectory to be deflected across the SLC and the ULC, and into the basin of the stable equilibrium, where it then remains in a quiescent non-spiking state. This is the mechanism by which the activity of the neuron terminates. (Of course panel B is a projection of the full dynamics onto the $V$-$n$ plane, so an apparent crossing in this plane does not necessarily correspond to a true crossing in the complete four-dimensional phase space. But the neuron's subsequent resting behavior confirms our interpretation.)

We performed a similar analysis for the electrical and inhibitory cases (not shown), and found that deflections of the trajectory due to synaptic currents were smaller and did not occur near the lower part, where the spiking state is vulnerable because the SLC and the basin boundary are extremely close together. Thus, spike termination is not seen in those cases.

The analysis above examines the termination effect in a single neuron. In a network, the size of the third phase depends on how many inputs are received by each neuron, and how they are temporally integrated. For network-level spike termination, each neuron, notably the one with the fewest connections, must receive a synaptic current with an above-threshold third phase. This is most effective with high synchronization.

\begin{figure}
\begin{centering}
\includegraphics[width=10cm]{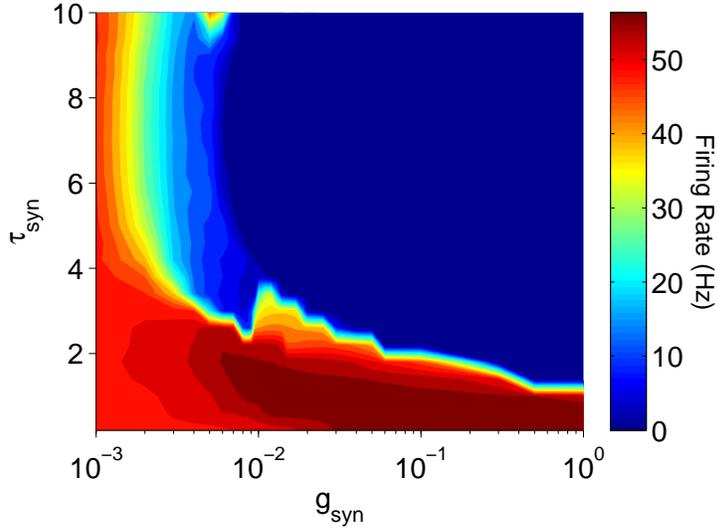}
\end{centering}
\caption{\setstretch{1.0}\label{contour_tg} The population mean firing rate as a function of the coupling strength $g_{syn}$ and the synaptic time constant $\tau_{syn}$. Dark blue denotes zero firing rate, and hence network spike termination.}
\end{figure}

Synaptic parameters such as the maximum synaptic conductance $g_{syn}$ and the time constant $\tau_{syn}$ influence the integration of synaptic inputs, and hence strongly affect the amplitude and duration of this third phase. Large values of $g_{syn}$ lead to more postsynaptic current, and large values of $\tau_{syn}$ lead to slower decay of these currents. Together, these effects lead to an increased amplitude and duration of the third phase of the synaptic current profile, and hence increase the probability for network spike termination to occur. This reasoning is confirmed in Figure \ref{contour_tg}, which shows the network population mean firing rate versus these two parameters.
We see that spike termination in the whole population occurs, in general, for sufficiently large values of both $g_{syn}$ and $\tau_{syn}.$

For small values of $\tau_{syn}$ (e.g., below $\sim 2$ ms), the synapses are fast in the sense that the synaptic variable decays away very quickly. Thus to achieve a large postsynaptic current, inputs must arrive in a highly synchronized fashion. Interestingly, we find that for values of $\tau_{syn}$ below $\sim 2$ ms, the mean network firing rate is mostly independent of the conductance $g_{syn}$. Furthermore, the network firing rate approaches that of an independent neuron at our parameters, indicating that almost all neurons are firing. Thus even with strong coupling, the inputs arrive scattered temporally, and the resulting very brief postsynaptic currents do not integrate together sufficiently to lead to spike termination.

On the other hand, for small values of $g_{syn}$ (e.g., below $\sim 2 \times 10^{-3}$), the coupling is so weak that the neurons essentially do not influence each other, and they fire at the rate of independent neurons. The exception is if $\tau_{syn} $  is large, so that the synaptic variable decays slowly enough that the inputs can sum significantly. We see that as $g_{syn}$ increases with $\tau_{syn} > 3$ or so, the network spiking rate falls gradually. This presumably reflects spike termination occurring first in highly-connected neurons, and then progressively more in less-connected neurons as the conductance increases.

We now consider the effect of channel noise on the spike termination phenomenon. Recall that in our implementation, channel noise is inversely related to membrane area. In all the results reported above, a fixed membrane area $A=10^5 \mu m^2$ was used. The noise present in this case can be considered to be small, as the results do not differ significantly from the zero-noise case.

\begin{figure}
\begin{centering}
\includegraphics[scale=.65]{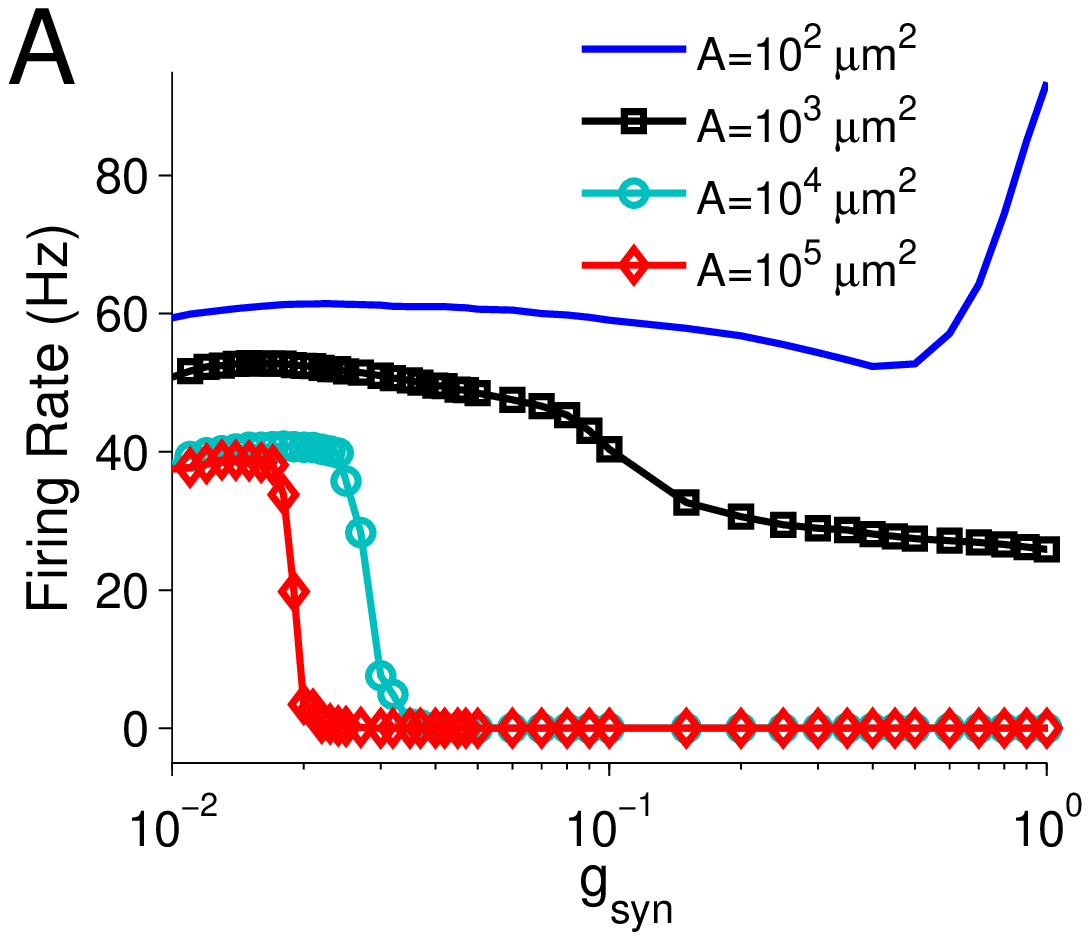}
 \hspace{3ex}
\includegraphics[scale=.65]{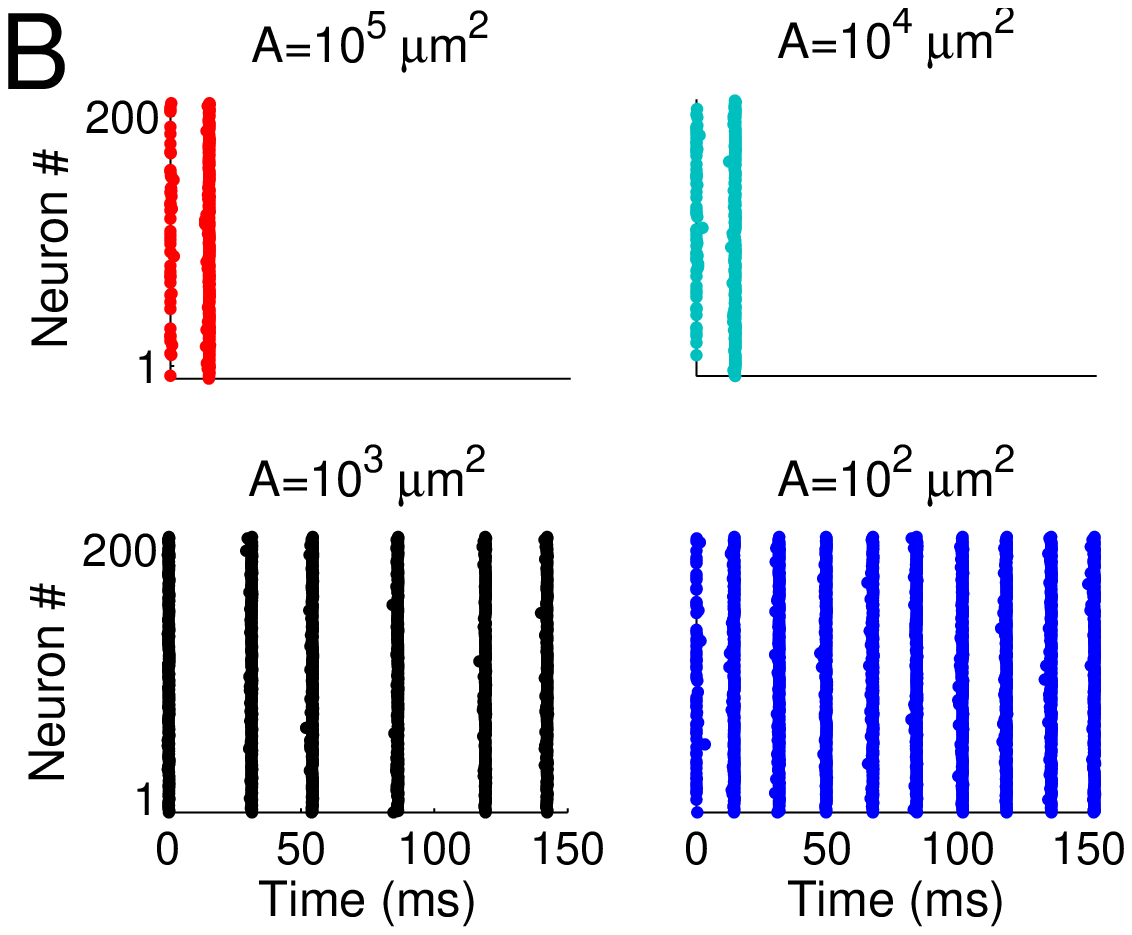}\\
 \vspace{.5cm}
\includegraphics[scale=.65]{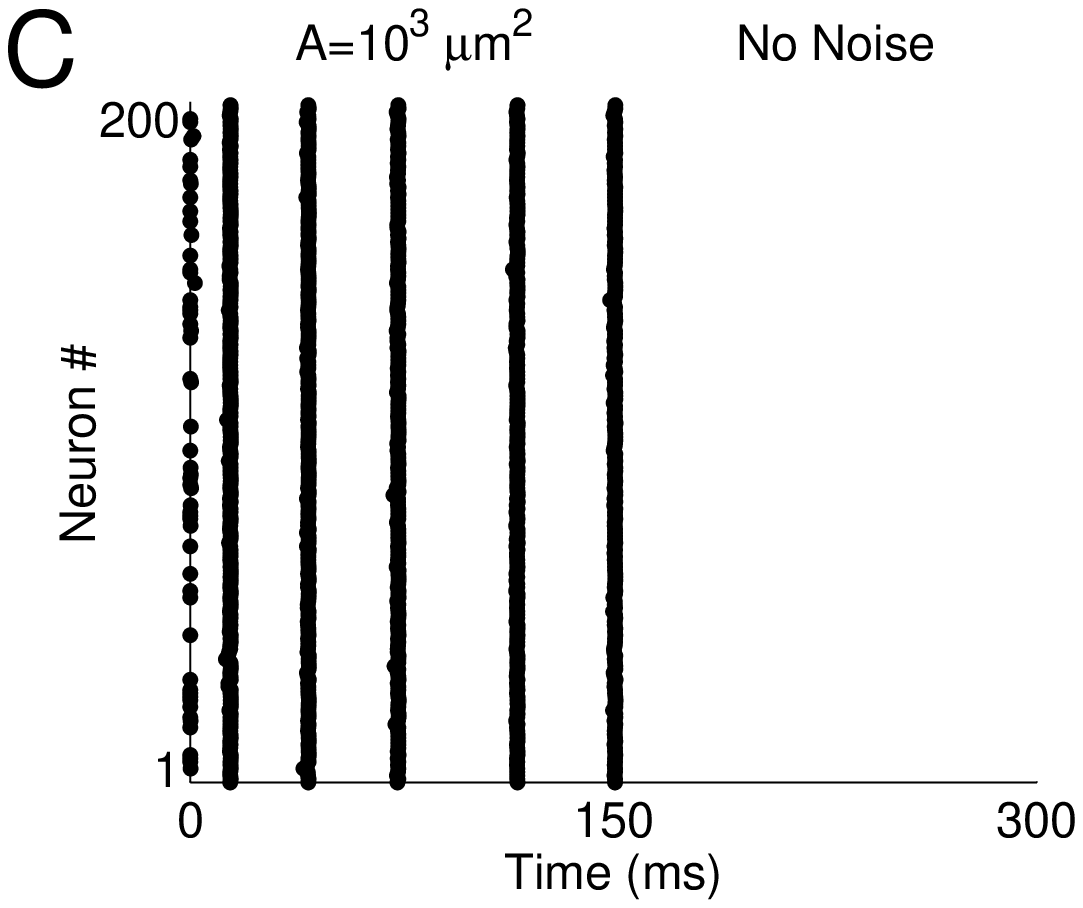}
\hspace{3ex}
\includegraphics[scale=.65]{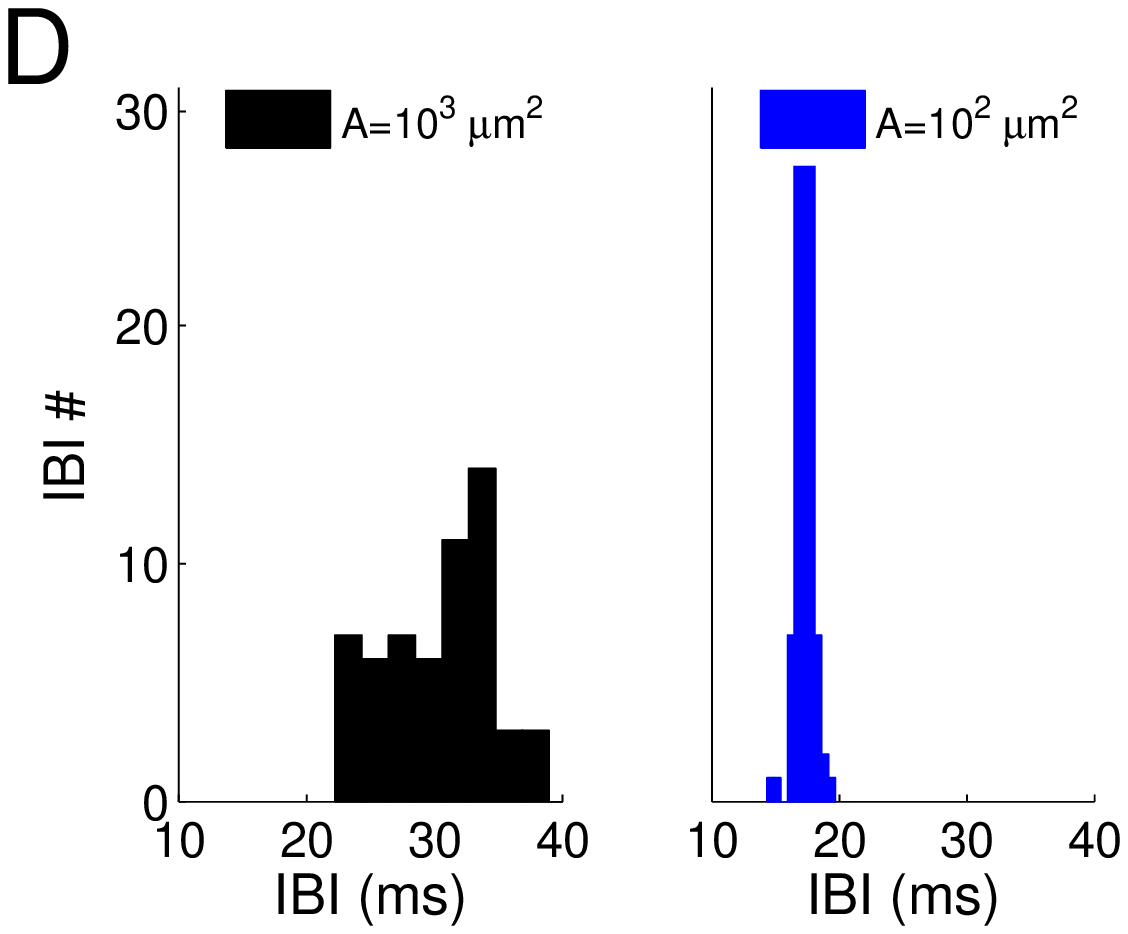}
\end{centering}
\caption{\setstretch{1.0}\label{area_k20}
The influence of channel noise on the spike termination phenomenon in excitatory networks. (A) Population mean firing rate versus $g_{syn}$ for different levels of channel noise parameterized by A. (B) Raster plots of population spiking activity for the cases shown in (A), with $g_{syn}=0.15$. (C) A raster plot for confirmation of spike termination occurrance when noise is removed from neural dynamics. Here, channel noise is turned off near $t=150 $ms. (D) Inter-burst interval (IBI) distributions for two values of $A$ as shown.}
\end{figure}

Fig.~\ref{area_k20}A shows the population mean firing rate for the excitatory network as a function of $g_{syn}$, for several (smaller) values of the membrane area. Note that the curve for $A=10^5 \mu m^2$ reproduces the curve for the excitatory network shown in Fig.~\ref{scats}A. As the membrane area decreases, the noise amplitude increases, and we see several effects. Initially, the onset of network spike termination simply shifts to higher values of $g_{syn}$. This is seen in the $A=10^5$ and $10^4 \mu m^2$ curves, which differ within the (approximate) range $g_{syn} \in \left[ .01, .03 \right]$. The difference observed within this interval is due to the desynchronizing effect of the noise at $A=10^4 \mu m^2$, which prevents the occurrence of a single strongly synchronous population event that causes spike termination (as is seen to occur in the $A=10^5 \mu m^2$ case). Second, the curves generally shift upwards, reflecting a higher network mean firing rate for larger amplitudes of noise. Finally, we see a sharp increase in the firing rate for large coupling strengths at the highest noise amplitude.

We looked more carefully at the network behavior for $g_{syn}=0.15$. Fig.~\ref{area_k20}B shows raster plots of the activity corresponding to the four curves in panel A. The first two plots ($A=10^5$ and $10^4 \mu m^2$) show that spiking activity terminates after an immediate strongly-synchronous population spike. For $A=10^3 \mu m^2$, we see similar large synchronous population spiking events, but these recur irregularly. Each of these population spikes in effect causes network spike termination (in the sense that all neurons are kicked off their spiking limit cycle as in Fig.~\ref{phaseplot}). But after a delay, a new population event is triggered by a random noise fluctuation in a highly-connected neuron. This neuron spikes, and very quickly thereafter the entire network is recruited. Thus the entire network behaves as a randomly-driven excitable system. This interpretation is confirmed in Fig.~\ref{area_k20}C, which shows a raster plot for this situation, but with the noise turned off after the last population spike near $t=150 $ms. Network activity ceases thereafter, so spike termination occurred. In addition, we see in the left panel of Fig.~\ref{area_k20}D that the corresponding inter-burst interval distribution reflects the randomness of the population spikes. For $A=10^2 \mu m^2$, corresponding to the largest noise amplitude, the population events occur more frequently and more regularly. Thus with more noise, the distribution of inter-burst intervals is narrower and is shifted towards smaller values (Fig.~\ref{area_k20}D, right panel), again supporting this probabilistic interpretation. We conclude that the spike termination phenomenon is severely affected by channel noise.

\section{Conclusions}

In this work, we studied the population behavior of scale-free networks of bistable Hodgkin-Huxley-type model neurons, with the neurons being connected by gap junction electrical connections, inhibitory synaptic connections, or excitatory synaptic connections. We observed the emergence of different patterns of synchronization among neurons and found an interesting behavior which only occurred in our networks with excitatory synapses: after a self-organized, strongly synchronous population spike, activity in the network ceased.

Examining the cause of this spike termination effect, we found that at the network level, a highly-synchronous population spiking event is necessary, but not sufficient. All three of our networks exhibited strong synchrony with appropriately strong coupling, but spike termination only occurred in the excitatory network. What distinguishes the three cases at the cellular level is the shape of the excitatory postsynaptic current (EPSC) received by the individual neurons during that synchronous event. In the excitatory network, this EPSC has a triphasic structure consisting of a very rapid depolarization followed by an immediate hyperpolarization, and finally a slow depolarization with a gradual decay. This triphasic EPSC profile arises from the competition between the postsynaptic membrane voltage and the excitatory synaptic reversal potential, and it depends strongly on the level of synchronization in the excitatory population.

We showed that the first two phases of this EPSC play no role in the spike termination effect. We further showed that termination occurs due to the third phase. If the third phase has sufficient amplitude and appropriate timing, spike termination occurs in our excitatory networks. The current amplitude depends on the integration of the arriving action potentials, to which network synchrony, the synaptic time constant, and the strength of synapses are important contributors. We found only a weak dependence on the latter two parameters, and that termination fails only when one of them is very small. Thus, we expect SIST to occur for typical synaptic and coupling parameters, and therefore may potentially be relevant in many real-world phenomena such as epileptic seizure termination and control of persistent sustained activity for working memory.

We developed an understanding of how the third phase of the EPSC leads to the termination of spiking in an individual neuron from a dynamical systems perspective. In the absence of synaptic inputs, the trajectory of a single regularly-spiking neuron follows a limit cycle that exists in the full state space. However, a neuron within a network receives inputs from other spiking neurons that perturb its trajectory away from the spiking limit cycle. Usually these perturbations cause only minor changes to the action potential's shape and/or timing. If a large and synchronous network spiking event occurs, then the integrated inputs to the neuron can lead to a special EPSP. If, in addition, the necessary dynamical structure exists, such an EPSP can push the neuron's trajectory into the basin of a co-existing resting state equilibrium. If, finally, this happens to all the neurons at approximately the same time, then the network activity ceases. This is what we observed in our excitatory networks. We also observed that the perturbations from the PSCs in the inhibitory and electrical networks are different, and do not lead to this effect.

We conclude that the necessary components for the occurrence of SIST are (1) a dynamical structure in the individual neurons containing bistability between a stable limit cycle and a stable fixed point, along with a region where the boundary between the basins of these attractors is relatively close to the limit cycle, and (2) a mechanism that gives rise to strong network synchronization.

\textcolor{red}{The} dynamical structure described in the first necessary component occurs in the Hodgkin-Huxley model neuron. We confirmed that SIST also occurs in excitatory networks of Morris-Lecar neurons in which the model parameters were chosen to exhibit the necessary dynamical features (data not shown). Accordingly, we expect SIST to occur in networks of other model neurons that can have similar structures, such as the Hindmarsh-Rose model \cite{HosakaSakai}, if the parameters are appropriately tuned. Furthermore, we found that SIST in our network was robust to heterogeneity in the excitability parameter $I_0$, as long as most neurons remained in the bistable region so that the required dynamical structure remained relatively intact (data not shown).

The nature of the network structure is important for the second necessary component (i.e., strong population synchronization) to arise. We chose to use scale-free network topology since it has been observed in the connectivity of functional brain regions via neuroimaging and electrophysiological studies, and in connectome data extracted from different diffusion magnetic resonance imaging (MRI) techniques \cite{chialvoprl, Li2010, Thivierge2014, Ciuciu2014}. Scale-free networks are topologically heterogeneous, yet can still give rise to large-scale synchronization as is required for the SIST mechanism described here. More homogeneous network structures (such as regular, small-world, all-to-all, random, etc.) also exhibit synchronization with sufficiently strong coupling strengths, so we expect that SIST could occur in these network structures as well.

This mechanism of spike termination is different from other mechanisms of oscillator quenching that have been discussed in the nonlinear dynamics literature, such as amplitude death and oscillator death. The latter two phenomena arise when the network coupling causes either the stabilization of a quiescent {\emph{unstable}} equilibrium, or the creation of {\emph{new}} stable quiescent state \cite{Saxena2012, Koseska2013}. Our situation is much simpler in the sense that a stable quiescent state already exists in the individual oscillators themselves. In addition, Ermentrout \cite{Bard} found that sufficient gap junction conductivity destroys persistent states of asynchronous activity in a network of conductance-based model neurons with excitatory synaptic coupling. Similar to our case, he found that population activity ceased following the development of a strongly synchronous state. However, this study considered Type-I neurons, for which the onset of spiking occurs via SNIC (saddle-node on an invariant circle) bifurcation. Thus, there is no bistability at the level of the individual neuron, as in our case. Instead, the mechanism for activity termination was found to be a homoclinic bifurcation in the dynamics of the network as a whole.

Returning to our work, we found that channel noise strongly influences SIST in such a way that above a certain amplitude, the absolute termination of network activity can be prevented. Although our results indicate that the SIST mechanism remains present even with large noise, the noise itself can stochastically trigger neurons to fire after termination has occurred. Subsequently, a new strongly-synchronized population spiking event emerges, and the activity terminates again. This noise-induced effect indicates that in essence, the entire population acts as a single excitable system which terminates and reignites irregularly in time. This might have implications for the emergence of several noise-induced phenomena in neural populations which are collectively excitable.

The present work can be extended to the study of networks with combinations of excitatory and inhibitory synapses present, with or without gap junctions, and perhaps with different and more realistic connection topologies. Another possibility would be to study the effects of using a different form of synapse. However, more realistic synapse models generally give rise to alpha-function-type PSCs that are combinations of exponential functions. We conjecture that since the third phase of the EPSCs that we considered here have qualitatively the same shape, spike termination may arise in such networks via the same mechanism.

Finally, our study could potentially be relevant for the understanding of sudden unexpected death in epilepsy (SUDEP), a rare phenomenon observed in people suffering from uncontrolled epileptic seizures \cite{Nashef2012}. Since epileptic seizures have their origin in an abnormal functioning of brain areas in which  there is at some time a strong level of synchronization, this could induce a sudden collective spike termination, as our study suggests. Such an event could subsequently propagate to other regions of the brain, possibly inducing the cessation of a critical physiological function and its fatal consequence.

\begin{acknowledgments}
 JJT acknowledges financial support from the Spanish Ministry of Science and the \textquotedblleft Agencia Espa{\~n}ola de
Investigaci{\'o}n (AEI)\textquotedblright$\,$  under grant FIS2017-84256-P (FEDER funds). We sincerely thank Sukriye Nihal Agaoglu for valuable discussions. We also gratefully acknowledge the anonymous reviewers for providing useful comments and suggestions, which greatly improved our paper. 
\end{acknowledgments}

\end{document}